\documentclass[preprint]{aastex}
\begin{document}
\received{ }
\accepted{  }
\journalid{ }{ }
\slugcomment{   }

\lefthead{Barrado y Navascue\'es et al. }
\righthead{Low Mass Members  of IC23291}

\title{Very Low Mass Stars and Brown Dwarfs of the Young  
Open Cluster IC2391\footnote{Based on observations obtained at
the Cerro Tololo Inter-American Observatory.} }

\author{David Barrado y Navascu\'es\altaffilmark{2,3},
\affil{Max-Planck Institut f\"ur Astronomie. K\"onigstuhl 17,
Heidelberg, D-69117 Germany. barrado@pollux.ft.uam.es}
 John R. Stauffer\altaffilmark{4}
\affil{Harvard--Smithsonian Center for Astrophysics,
       60 Garden St., Cambridge, MA 02138, USA. jstauffer@cfa.harvard.edu}
 C\'esar Brice\~no\altaffilmark{2}
\affil{Centro de Investigaciones de Astronom{\'\i}a (CIDA)
   Apartado Postal 264
   M\'erida, 5101-A
   Venezuela. briceno@cida.ve}
 Brian Patten\altaffilmark{2}
\affil{Harvard--Smithsonian Center for Astrophysics,
       60 Garden St., Cambridge, MA 02138, USA. bpatten@cfa.harvard.edu}
 Nigel C. Hambly
\affil{Institute for Astronomy, University of Edinburgh, 
Royal Observatory, Blackford Hill, Edinburgh EH9 3HJ, Scotland, UK.
nch@roe.ac.uk}
\and
 Joseph D.  Adams
\affil{University of Massachusetts, Department of Physics and
Astronomy, Amherst, MA 01003, USA. adams@pegasus.astro.umass.edu}
}

\altaffiltext{2}{Visiting Astronomer, 
Cerro Tololo Inter-American Observatory. 
CTIO is operated by AURA, Inc.\ under contract
 to the National Science Foundation.}

\altaffiltext{3}{Present address: Departamento de F\'{\i}sica Te\'orica, C-XI.
Universidad Autonoma de Madrid, Cantoblanco, E-28049 Madrid, Spain }

\altaffiltext{4}{Present address: IPAC,  California Institute of Technology,
 Pasadena, CA 91125, USA}

\begin{abstract}

We have identified a large sample  of probable low mass members
of the young open cluster IC2391 based on optical (VRIZ) and 
Infrared (JHK$_s$) photometry. Our sample includes
50 probable members and 82 possible members, 
 both very low mass
stars and brown dwarfs.
We also provide accurate positions for these stars and brown dwarf
candidates  derived from red UK 
Schmidt plates measured using the microdensitometer SuperCOSMOS. 
Assuming an age of 53 Myr, we estimate that we have reached 
a mass of 0.025 M$_\odot$,
 if the identified objects are indeed members of IC2391.

\end{abstract}

\keywords{stars: low mass, brown dwarfs, open clusters
and associations: IC2391}

\section{Introduction}

IC2391 is one of  the youngest and nearest open clusters.
The  {\it Hipparcos} distance modulus for the cluster  is (m-M)$_o$$\sim$5.82.
As a compromise between various determinations
of the distance modulus and {\it Hipparcos} parallaxes
(Becker \& Fenkart 1971; Lyng\aa{ } 1987; van Leeuwen 1999;
 Robichon et al.  1999),  we adopt (m-M)$_0$=5.95$\pm$0.1,
corresponding to a distance of 155 pc. 
The mean reddening towards the cluster is small,
E(B-V)=0.06 (Patten \& Simon 1996).
Prior to our lithium depletion age determination, the most 
commonly quoted age estimate for IC2391 was $\sim$35 Myr 
(Mermilliod 1981). In Barrado y Navascu\'es, Stauffer \& Patten (1999)
we derived a new age estimate of $53 \pm 5$ Myr.

Because of its proximity and youth, IC2391  has been the target of a number 
of recent studies. Rotational velocities, lithium abundances and H$\alpha$
data for G, K and early M  dwarf members can be found in
Stauffer et al. (1989, 1997). X-ray fluxes,
rotational periods and low resolution spectra for another set of probable
cluster members are provided by Patten \& Simon (1986) and Simon \& Patten 
(1998).
 All these data support the conclusion that IC2391
is  younger than the Alpha Per cluster ($\sim$85 Myr in the new
lithium age scale, Stauffer et al. 1999; 
Barrado y Navascu\'es, Stauffer, \& Bouvier 1999;
 Stauffer \& Barrado y Navascu\'es 2000)
but significantly older than 10 Myr (e.g., because the cluster
contains no classical TTauri stars).

For an age of 35 Myr, we expected the lithium depletion boundary for IC2391
to be at I$_C$$\le$15.2 (Baraffe 1998, private communication; 
Stauffer, Schultz, \& Kirkpatrick 1998). The ROSAT images of the cluster
only supplied candidate cluster members down to I$_C$$\sim$15.
No appropriate deep photographic plate sequences are available for
a proper motion survey to the desired depth. We therefore obtained
deep, multicolor photometric imaging of the cluster in order to produce
a list of candidate cluster members. Because of the low galactic latitude
 of the cluster ({\it b}$^{\sc II}$=-6.90), our derived
candidate list may be significantly contaminated by field stars. We therefore
defer extensive discussion of these candidates until detailed spectroscopic
follow-up is available.

In Section 2,  we  provide the details of our imaging program. The method
we used to identify candidate cluster members is outlined in Section 3.

\section{Observations and Data Reduction}

\subsection{The optical photometry and the initial selection of candidates}

 Our optical CCD photometric data (V,R$_C$,I$_C$,Z) 
were collected at the Cerro Tololo 
Inter-American observatory during four different observing runs:
January 22, 1998, at the 0.9m (RI$_C$ filters); 
 January 24--25, 1998, at the Blanco 4m telescope (RI$_C$ filters);
April 4, 1998, at the 1.5m (RI$_C$  filters); 
and January 4--6, 1999, again with the 0.9m telescope (VI$_C$Z filters).
 In the case of the CTIO 0.9m runs, we used the 
 Tek 2048 \#3 camera  (0.43 arcsec/pix),
yielding a field  of view   $\sim$14 arcmin on a side.
For the CTIO 1.5m telescope,
we used the CCD SITe $2048 \times 2048$ imager at the f/7.5 focus,
 yielding an image scale of 0.44"/pix and
a field of view of $14.8\times14.8$ arcmin. 
For the Blanco telescope campaign, we used the Big Throughput
Camera (BTC), a mosaic detector composed of four different
2048$^2$ pixel CCDs. Each CCD covers an area of 14.7$\times$14.7
arcmin, and the mosaic has a cross-shaped gap,  5.4 arcmin wide, 
 between the CCDs. 
The total projected area on the sky of the BTC Mosaic is roughly
0.25 sq. deg., with a scale of 0.43 arcsec/pixel.
The January 1998 and April 1998 runs were used to make sure we would be 
able to calibrate the BTC observations and to cover the gaps between 
the CCDs of the BTC mosaic. In total, we have covered an area close
to 2.5 sq. deg. (2 sq. deg. with the BTC).  Table 1 lists the 
coordinates of the center of each field, as well as the exposure times.
Figure 1 shows the location of the observed area, together 
with our very low mass (VLM) stars and brown dwarf (BD) candidates
(cross symbols).

The CCD images
were bias-subtracted and flat-fielded using standard data reduction
techniques and tools within IRAF\footnote{IRAF is distributed by the 
National Optical Astronomy Observatories, which is operated by the
Association of Universities for Research in Astronomy, Inc., under
contract with the National Science Foundation.}.  The APPHOT package was 
used to extract instrumental magnitudes for the objects of interest
in each CCD field.

\subsection{The calibration of the data}
All our 0.9m and 1.5m data were obtained under photometric 
conditions.  Unfortunately, 
this was not the case for the CTIO 4m run (BTC data). Therefore, we  used
the first two sets of data from the 0.9m and 1.5m telescopes
 to calibrate the BTC data,  corresponding to the (RI)$_C$ filters.
This calibration included several steps. First, we extracted
the instrumental magnitudes for all the CCDs using small apertures
($\sim$2 pixels, equivalent to $\sim$0.9 arcsec). Then, for each image,
 we derived an aperture correction ($\sim$0.12 mag).
 Using this method, we minimized the 
photometric errors, since we did not include
flux from the nearby sky. 
 Then, for the 0.9m and 1.5m data, 
we corrected for the airmass. We assumed the
standard CTIO extinction coefficient (0.08 mag/airmass and 0.06
mag/airmass for the R$_C$ and I$_C$ filters, respectively).
Standard stars from Landolt (1992) were  observed in all cases, 
including the fields SA98, SA104, PG1047 and PG1323.
 They were used to calibrate
the data into the Cousins system.
 Finally, the calibrated photometry from
the 0.9m and 1.5m runs was compared with the instrumental magnitudes
of the BTC campaign.  This provided
a zero point shift for each filter  from which we
computed calibrated magnitudes for the BTC survey.
 We did not find any significant color term
between the three cameras/filter systems.

Table 1 lists the limiting magnitudes for each field in the 
R$_C$ and I$_C$ filters. The internal errors should be better than 
 0.15 magnitudes at the 
limiting magnitudes;
and the photon statistical errors are 
$\sim$0.01 mag for R$_C$$\sim$17 and I$_C$$\sim$16 for the 0.9m and 1.5m data,
and 0.01 for R$_C$$\sim$19 and I$_C$$\sim$17.5 for the BTC data.
We have estimated the completeness limits from the histogram
Log N$_{\rm detections}$ versus magnitude.
These limits are defined by the magnitude where the histogram deviates 
from a straight line (see Wainscoat et al. 1992 and Santiago et al. 1996).
 In the case of the 0.9m and 1.5m data, they are
R$_C$$\sim$18.5 and I$_C$$\sim$18.0 mag. For the BTC data, we have
reached  R$_C$$\sim$20.5 and I$_C$$\sim$20 mag.

In order to check the external accuracy of our optical data, we have
carried out different comparisons. 
As stated before, we calibrated the BTC data using 2 different 
campaigns (0.9m and 1.5m telescopes)
 which took place immediately before and after that run.
The comparison between  these two runs gives dispersions
of $\sigma$(R)=0.08, $\sigma$(I)=0.07 and $\sigma$(R-I)$_C$=0.01. 
The calibration of the BTC photometry 
has dispersions of 0.03 and 0.04 magnitudes for the R$_C$ and I$_C$ filters,
 respectively (rms scatter of the difference between the raw 
and calibrated magnitudes of  the standard stars).
Finally, the optical photometry
for overlapping fields of the BTC survey agrees within 0.06 and 0.04
magnitudes (one sigma) for the R$_C$ and I$_C$ filters, after the calibration.

For the last observing run (January 1999, CTIO 0.9m),  we used the VI$_C$Z
filters and the photometry   was calibrated
independently.  The V-band instrumental magnitudes were corrected
for the effects of atmospheric extinction and were placed on the 
standard system (Johnson-Kron-Cousins) using observations of 
photometric standard stars (Landolt 1992) and previously determined
(R-I)$_C$ colors for our program objects.  
 Because the I$_C$ and Z CCD images were 
taken in consecutive pairs (i.e., at the same airmass) and because
at that time there were  no well-established standard stars for Z filter 
photometry in the literature, we did not perform an absolute 
calibration of the I$_C$ and Z data.  Instead, an (I-Z)$_{CTIO}$
color index was calculated using just the $I$ and $Z$ instrumental 
magnitudes.  As in Zapatero Osorio et al. (1999), we set I$_C$=Z
for those standard stars observed with (R-I)$\sim$0 in order 
to determine a zero-point correction for our (I-Z)$_{CTIO}$
color index.

  The internal errors of our V-band photometry are estimated to
range from $\sim$0.02 mag for V$\sim$17 to $\sim$0.15 mag for
 V$\sim$21.  The external errors on the V magnitudes may
be larger since our typical program stars are much redder than
any of the standard stars used to determine the transformation
to the standard system.  The uncertainties in the (I-Z)$_{CTIO}$
color are estimated  from the errors in the I$_C$ and Z magnitudes
as supplied by the PHOT routine in IRAF.
They are computed as 
$\Delta$(Z-I)$^2$=$\Delta$Z$^2$$+$$\Delta$I$^2$.
  These should be on
the order of 0.02 mag or less for the majority of the objects in
our IC 2391 sample (for I$_C$$\sim$17 mag).

\subsection{Infrared photometry}

The infrared data were obtained by the 2MASS survey 
(Skrutskie et al. 1997)
at the CTIO facility during November 21-25 and November 28-29, 1998
and processed at IPAC in January, 1999. All 2MASS scans used 
for this analysis were observed under photometric conditions. 
 Based on 
photometric error and total sources detected with magnitude we
estimate that these 2MASS data still reach
$\ge 0.99$ completeness and SNR $\ge 10$ at $J=15.8$, $H=15.1$, 
and $Ks=14.3$ despite any possible confusion noise in this field.

Both our optical data and the 2MASS survey have very accurate coordinates
(see next subsection).
Therefore, using the optical coordinates as input, we searched
in the 2MASS catalog for a 
IR counterpart, using a 3 arcsec radii for this search, much
larger than the errors in the coordinates.

\subsection{Coordinates}

To define astrometric solutions for the CCD frames we used
 secondary astrometric standards derived from United Kingdom 
Schmidt photographic plate  material
measured using the precision microdensitometer SuperCOSMOS 
(eg. Hambly et al. 1998). The global astrometric solution for 
the Schmidt plate  was derived using
the Tycho--ACT reference catalog (Urban, Corbin \& Wycoff 1998),
 and includes correction for non-linear systematic effects caused 
by the mechanical deformation of the plates during exposure 
(eg. Irwin et al. 1998). We used the "short red" survey plate R6843 
(epoch 1981.3, field number 165) for IC2391. These exposures,
taken at low galactic latitudes, are far less crowded than the sky limited
survey plates and reach R$\sim$20 (as opposed to R$\sim22$ for the deep
 survey plates). They are ideal for accurate astrometry of secondary 
standards as faint as R=20 which overlaps with unsaturated objects on
 the CCD frames. The RMS residual per ACT star in the global
 astrometric plate solution was $\sim0.1$ arcsec in both coordinates;
 we estimate that there will be no systematic errors
in the CCD astrometric solutions larger than this value.
 The   positions of our IC2391 candidates are listed in Table 2.

\section{Discussion}

\subsection{Initial selection of candidate members of IC2391}

The initial selection of candidate members of IC2391 
was carried out using the location of the detected stars on 
the I$_C$ versus (R-I)$_C$ color-magnitude diagram. 
Figure 2 displays  the data obtained
at the 4m CTIO/BTC telescope, where detections are shown as
dots. An empirical Zero Age Main Sequence (ZAMS), based on data
from Leggett (1992),  is included
as a solid line.   The ZAMS is plotted for our assumed
IC2391 distance modulus and a reddening of 
E(R-I)=0.007   [David:  is this E(R-I) compatible with the
E(B-V) mentioned earlier in the text?]
Completeness and limiting magnitudes are depicted as dashed lines.
We selected all the detected objects in a 
wide band well  above the ZAMS. 
  The lower envelope for the strip follows a 50 Myr isochrone
(D'Antona \& Mazzitelli 1997) and the upper envelope is
displaced 0.75 mag brighter to include equal mass binaries.
The boundaries of the strip were also adjusted to take 
into account the   photometric errors and errors in the
distance, age  and the reddening.
Hence, this band is  wider than the 0.75 mag.
Then, all candidates  were visually
inspected on the original images, in order to avoid the presence of
false detections, and  to verify the stellar-like shape of the detections 
 and the lack of nearby bright stars or cosmic rays which could modify the
photometry of the candidate.
Since there is no clear separation between the field stars and the
location of the cluster isochrone, we expect a strong contamination
by spurious members. Specifically,
those candidates located only slightly above the ZAMS
should be considered with some caution, since they
are well below the 50 Myr isochrone.
In total, we have selected 206 candidate members, 94 from the
BTC survey and another 112 from the 0.9m and 1.5m data.
Figure 3 shows all our candidate members as circles. Solid circles
indicate the position of the candidates whose membership has
been confirmed  spectroscopically, via radial velocity,
spectral type and several spectroscopic features (H$\alpha$, 
NaI8200\AA). These results are discussed in
Barrado y Navascu\'es, Stauffer,  \& Patten  (1999), 
where we provide a lithium depletion boundary (see Barrado y Navascu\'es,
Stauffer \& Bouvier 1999)
 age for the cluster (53$\pm$5 Myr).
Figure 3 also displays the position of a 50 Myr isochrone (short-dashed
line), adapted from D'Antona \& Mazzitelli (1997). Based on  
our spectroscopically confirmed members and on Simon \& Patten (1996) data, 
we have created an empirical IC2391 isochrone, shown as a long-dashed line.

Table 2 lists the positions, optical and infrared 
photometry, the separation  between the optical detection and the IR source
 and the identification with stars from the literature, for  each candidate.

\subsection{Color-Magnitude and Color-Color Diagrams of IC2391}

The merger of our RI$_C$ data with the 2MASS  photometry, 
and the additional VI$_C$Z photometry collected in January 1999,
allows us to create a large database of optical-infrared 
broadband photometry in 7 different filters. Therefore,
 we have been able to construct several color-magnitude and
color-color diagrams. We have used these diagrams to estimate
the membership status of each candidate.

Figure 4a depicts the [V,(V-I$_C$)] diagram.
 The empirical main sequence (MS) for young disk 
stars by Leggett (1992) is included as a solid line. 
This MS was built for M0-M9 dwarfs, in the ranges
6.65 $\le$ M(I$_C$) $\le$ 14.67,
4.77 $\le$ M(K$_{cit}$) $\le$ 10.17,
0.75 $\le$ (R-I)$_C$ $\le$ 2.30, and
0.17 $\le$ (H-K) $\le$ 0.48.
Note that our 2MASS data were taken in K$_s$. However, 
there are  no important differences between  the K$_s$
and K$_{cit}$\
systems (see Persson et al 1998).
For comparison purposes, the shift of the photometry corresponding to an
interstellar absorption of A$_v$=2 is plotted as an arrow.
 Proposed members of IC2391 from the literature
(Patten and Simon 1996; Patten and Pavlovski 1999) appear
as crosses.
 Figure 4b displays the
I$_C$ magnitudes against the (I$_C$-K$_s$) color, 
whereas  Figure 4c shows the
K$_s$ magnitudes against the (J-K$_s$).
In all these figures,  solid triangles represent
initial candidate members whose membership has been rejected  based
on  Figures 4 and 5 (probable non-members),
 whereas open triangles  show the possible non--members.
Possible and probable members are represented by open and solid
circles, respectively (see next subsection).

Several color-color diagrams can be found in Figure 5.
Symbols are as in Figure 4. Panel a and b depict 
[(V-R$_C$),(R-I)$_C$],
and  [(I$_C$-K$_s$),(R-I)$_C$],
 respectively. 
It is clear from
examination of these diagrams that our original sample included both
stars that are plausible cluster members and objects that are
instead very likely to be heavily reddened, background stars.

\subsection{A final list of photometric candidate members}

In order to  remove the spurious members present in our 
  candidate member list, 
we have used several color-color diagrams  and 
color-magnitude diagrams. This selection was carried 
out in a hierarchical way, stressing the spectroscopic and IR data.
 The scheme we followed is described below:

\begin{enumerate}

\item A fraction of our candidates has intermediate 
resolution spectroscopy (Barrado y Navascu\'es et al. 1999), 
which indicates whether they are probable members or non-members.

\item We have removed the objects having a large interstellar reddening
(see Figure 4b,c and Figure 5b).
 They were classified as probable non-members.

\item Color-color diagrams  were used to remove additional  probable
non-members (Figures 5a,b).

\item  Objects fainter than I$_C$=17 mag, and without IR data (they are
too faint to be detected by the 2MASS survey) were classified as 
possible members, whereas stars brighter than this value, with no IR data
(they should have been detected by 2MASS) are listed as possible non-members.

\item Objects bluer than the Leggett's (1992) IR  main-sequence 
appear in Table 2 as possible members.  

\item Objects bluer than our empirical optical  isochrone
or  at the upper (bright) edge of our CM diagram selection strip
were also classified as possible members. 

\item Finally, the  objects that remain are considered to be
bona fide members of the 
cluster (``probable members").

\end{enumerate}

As a summary, we have classified
 our initial IC2391 candidate members in four
different categories:

\begin{itemize}

\item  Probable members.
Objects located in all  CM and CC diagrams
with positions which indicate membership. They are identified with the
flag ``{\it MEM}'' in the last column of Table 2 (50 objects, including 16 
having spectroscopy).
 
\item Possible members (82 objects).
 They appear
 identified with the flag ``{\it MEM}?'' in Table 2.
 
\item Possible non-members (10).
 (``{\it NM?}'' flag).
 
\item Probable non-members (64). 
All of them  are flagged with ``{\it NM}'' in  Table 2.

\end{itemize}
 
If we only consider our probable  members, we
 have detected objects $\sim$2 magnitudes fainter than the previous surveys
(Rolleston \& Byrne 1997; Simon \& Patten 1998; Patten \& Pavlovsky 1999).
Our faintest possible IC2391 member has I$_C$=20.9, which is 5 magnitudes
fainter than the least massive candidate discovered to date.
In total, we list 132 objects as probable or possible members
of IC2391.

We have  not  computed the luminosity function (LF) and
mass function (MF). Due to its
low galactic latitude  of the cluster ({\it b}$^{\sc II}$=-6.90),
even when using the probable/possible member list,
pollution by spurious members is likely to be significant.
Visual inspection of the I,(R-I)$_C$ color-magnitude
diagram (Figure 3) reveals two important characteristics:  the relative
high number of candidate members with I$_C$ in the range 19.5--20.9, and
the apparent gap just before this clustering occurs.
 only additional
photometry or spectroscopy would allow us to establish if these
 faint candidates are real. However, we expect a strong contamination
by field stars in  this range. If most of these objects are, indeed, not
members, the gap would be illusory and the low number of candidate
members below I$_C$=18  could be a consequence of our completeness
limit (dotted line in Figure 3). The question is still open,
until a follow-up spectroscopic study is carried out.
 
\subsection{The contamination by field stars}
 
 In a deep optical survey of the Pleiades, Bouvier et al. (1998)
estimated that the contamination due to field stars was 25\%. This
value has been confirmed by subsequent spectroscopic follow-up
of the Pleiades candidates. Since  IC2391 is closer to
the Galactic Plane than the Pleiades ({\it b}$^{\sc II}$=-6.90
and  {\it b}$^{\sc II}$=-23.52, respectively), the contamination should
be much stronger. In fact, Figure 2 of Bouvier et al. (1998) shows a
clear   discontinuity between field stars and
the Pleiades population.  This is not the case for IC2391,
whose color-magnitude digram (Figure 3) depicts  a smooth transition between
the field and  the locus of the main sequence of the cluster,
indicating that contamination by spurious members should
be stronger than in the case of the Pleiades.
To estimate the degree of contamination by field stars, we have constructed
histograms of the number of stars per magnitude bin at various (R-I)$_C$
color intervals. In Figure 6 we show the histogram for the
(R-I)$_C$=1.9-2.0
range, representative of the distribution of stars down to approximately
the completeness limit of the BTC data. The locus of the cluster is
indicated. Similar diagrams for adjoining color ranges provide analogous
results. The last bin before the cluster
contains 3 stars, whereas the brightest bin, well above the
main sequence for equal mass binary  cluster stars (0.75 brighter than the
single star MS), has 5 stars. The bins in between have 17 stars.
Assuming  an average contamination of 4 stars per bin, the pollution rate
would be $\sim$50\% for the initial list of IC2391 candidates (the 
RI$_C$ survey).
Our optical and infrared color-color and color-magnitude diagrams
have allowed us to remove about 33\% of the candidates from  the initial
206 objects. Therefore,  we estimate that 
objects cataloged as  members in Table 2 (probable and possible members)
still have a
$\sim$25\% or greater probability of being spurious. 
 
The faintest objects,
lacking VZJHK data, may be more strongly polluted by field stars.

\subsection{Brown Dwarfs in the Cluster}

 Based on the position of the lithium depletion boundary in
members of IC2391, 
Barrado y Navascu\'es, Stauffer,  \& Patten (1999)
have recently determined an age for the cluster of 
53$\pm$5 Myr. For this age, the 
border between the stellar and substellar regimes would be at 
M(I$_C$)=11.06 (Baraffe  2000, private communication).
 Taking into account the  IC2391 distance and
reddening, this leads to I$_C$=17.03. Therefore, all our IC2391
candidates fainter than that value should be brown dwarfs
if they indeed belong to the cluster.
  Our list of final candidates contains ten objects which have been
cataloged as probable and possible  members based on both optical and 
infrared data and  which have I$_C$ in the range 17.06--17.62.
Using the Baraffe (2000) models, their mass range is
0.070--0.055 M$_\odot$. Actually, two of these objects (CTIO-061 and
CTIO-113) have been observed using   low S/N, intermediate resolution
spectroscopy (Barrado y Navascu\'es, Stauffer,  \& Patten 1999).
 Several spectral 
characteristics, such as H$\alpha$ and NaI8200\AA{ } equivalent widths, 
spectral type, rough radial velocities, indicate that they are real 
members of the cluster and, therefore, brown dwarfs.

Another 50 objects  in Table 2 are fainter than than the I$_C$ magnitude of the
stellar/substellar limit and lack infrared data.
All of them  are listed as possible members in Table 2, 
with magnitudes down to I$_C$=20.9, M(I$_C$)=14.93, and M $\ge$ 0.025  
M$_{\odot}$.

\subsection{Comparison with previous surveys and  other clusters}

A comparison between our final candidate members  (circles) 
and previous surveys (triangles)
of IC2391 is shown in Figure 7a. Probable candidate members 
are displayed with solid circles, whereas open circles represent
objects which  do not have infrared counterparts (normally, because
they are too faint to be detected by 2MASS). The stars from Simon 
\& Patten (1996)  are shown as solid triangle, and the objects
from Patten \& Pavlovsky (1999) appear as open triangles.
Clearly, all these samples merge smoothly, describing a good 
cluster isochrone. Most of  our probable candidate members are above
 the empirical  isochrone,
which was obtained based on confirmed members of the cluster (Figure 3).
 In fact, they are located in a  wide band,  slightly larger than
$\sim$0.75 mag (the maximum  shift due to binarity). 

We have compared our optical data of IC2391 candidates with 
data from two other  very well known clusters. Figure 7b contains
photometry from the Pleiades (asterisks), alpha Per
(plus symbols) and IC2391 (solid and open circles for probable and possible
members).
These open clusters have lithium ages of 125, 85 and 53 Myr, respectively
(Stauffer,  Schultz, \& Kirkpatrick 1998; Stauffer et al. 1999;
Barrado y Navascu\'es, Stauffer, \& Patten 1999).
The large  width  of the IC2391 main sequence, which includes
probable and possible members, can be appreciated. 
However, this scatter is greatly reduced if we only take into account
probable members. Then, the scatter is 
 similar to the value present in the other two 
clusters. This is consistent with our previous estimate of the 
contamination for the possible members which are located
at the blue side of our empirical IC2391 isochrone.
Only further spectroscopy (or, eventually, proper
 motions) will allow us to verify the membership status of these objects.
The comparison between the MS lower envelopes  of these two clusters with the 
lower MS of IC2391 probable members suggests that, indeed, IC2391 is
slightly younger than Alpha Per, and that both of them are
considerably younger than the Pleiades.

\section{Summary}

Using  optical and infrared data, and based on the location
on  CM and CC diagrams, we have identified a large sample of 
very low mass stars and brown dwarf candidates of the young cluster
IC2391. Accurate coordinates, derived from the SuperCOSMOS
microdensitometer, are provided for all of them.
We have established a total of 50 probable cluster members,
based on their position in the cluster loci in all CC and CM diagrams.
 Another 82 objects have been
cataloged by us as possible members.
We have identified two candidate sub-stellar
IC2391 members for which we have a full set of
multicolor photometry, and whose location in all
the CM and CC diagrams supports cluster membership.
We have an additional 50 candidate
substellar mass members in IC2391, but because we have
less or poorer data for these objects we expect
that many of them will instead be low mass field stars.

\acknowledgements

 DBN thanks the  {\it ``Instituto Astrof\'{\i}sico de Canarias''}
 (Spain) and the
{\it ``Deutsche Forschungsgemeinschaft''} (Germany) for  
their fellowship. JRS acknowledges support from NASA Grant
NAGW-2698 and 3690.
This work has been partially suported by 
Spanish {\it ``Plan Nacional del Espacio''}, under grant ESP98--1339-CO2.
This publication makes use of data products from the Two Micron All Sky
Survey, which is a joint project of the University of Massachusetts and
the Infrared Processing and Analysis Center, funded by the National
Aeronautics and Space Administration and the National Science Foundation.


\newpage

\begin{center}
{\sc Figure Captions}
\end{center}

\figcaption{Positions of our candidate members of IC2391 (plus
symbols). The brightest stars in the field are represented
as  four-pointed stars.
\label{fig1}}

\figcaption{All the photometry extracted from the 4m CTIO/BTC
survey. The solid line indicates the positions of a ZAMS, whereas the 
long-dashed lines show the location of the detection and completeness
limits. The completeness
limit of the BTC survey is indicated by a dotted line.
\label{fig2}}

\figcaption{Initial selection of candidates of IC2391 (open circles)
based on RI$_C$ data.
Confirmed members (Barrado y Navascu\'es et al. 1999)
 are displayed as solid circles. The solid, long-dashed and short dashed
 lines indicate the positions of a ZAMS, an empirical IC2391 isochrone
and a 50 Myr isochrone (D'Antona \& Mazzitelli 1997).
\label{fig3}}

\figcaption{
 Color-magnitude diagrams  of IC2391. Solid circles represent 
the probable members of the cluster, whereas open circles correspond
to possible members.
Open and solid triangles  are objects initially selected as members,
whose membership has been rejected based of these CC and CM diagrams.
Data from previous surveys appear as  crosses.
The solid line represents the locus of an empirical ZAMS
(Leggett 1992), whereas the long-dashed line (panel a and b) corresponds
to an empirical IC2391 isochrone. Panel a also includes a 50 Myr
isochrone by D'Antona \& Mazzitelli (1997).
\label{fig4}}

\figcaption{Color-color diagrams of IC2391. Symbols as in Figure 4.
\label{fig5}}

\figcaption{Number  of stars with 1.9$\le$(R-I)$_C$$<$2.0
against the I$_C$ magnitude. The location of IC2391 is indicated.
\label{fig6}}

\figcaption{Absolute I$_C$ magnitude against the unreddened 
(R-I)$_C$ color index.
Comparison with data from previous searches of
members of IC2391 (panel a) and members of other clusters (panel b).
An empirical ZAMS and an empirical IC2391 isochrone are represented
as solir and dashed lines, respectively.
IC2391 data comes from this survey, Simon \& Patten (1996) and
Patten \& Simon (1996). Alpha Per data were selected from 
Prosser (1992, 1994) and Stauffer et al. (1999). Finally, 
Pleiades data comes from Bouvier et al. (1998).
\label{fig7}}

\newpage

\setcounter{figure}{0}
\begin{figure*}
\vspace{18cm}
\includegraphics{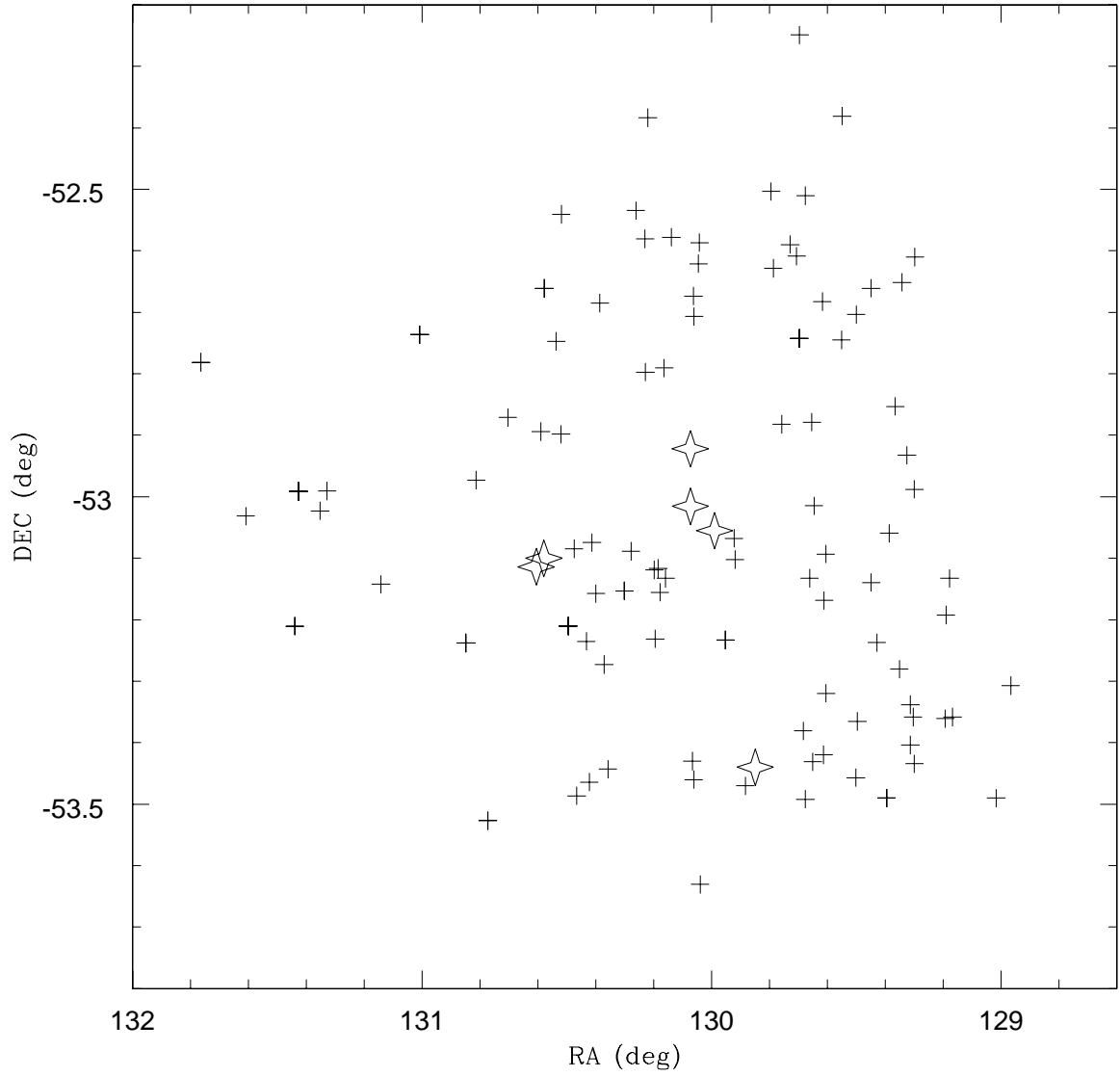}
\caption{Positions of our candidate members of IC2391}
\end{figure*}

\setcounter{figure}{1}
\begin{figure*}
\vspace{18cm}
\includegraphics{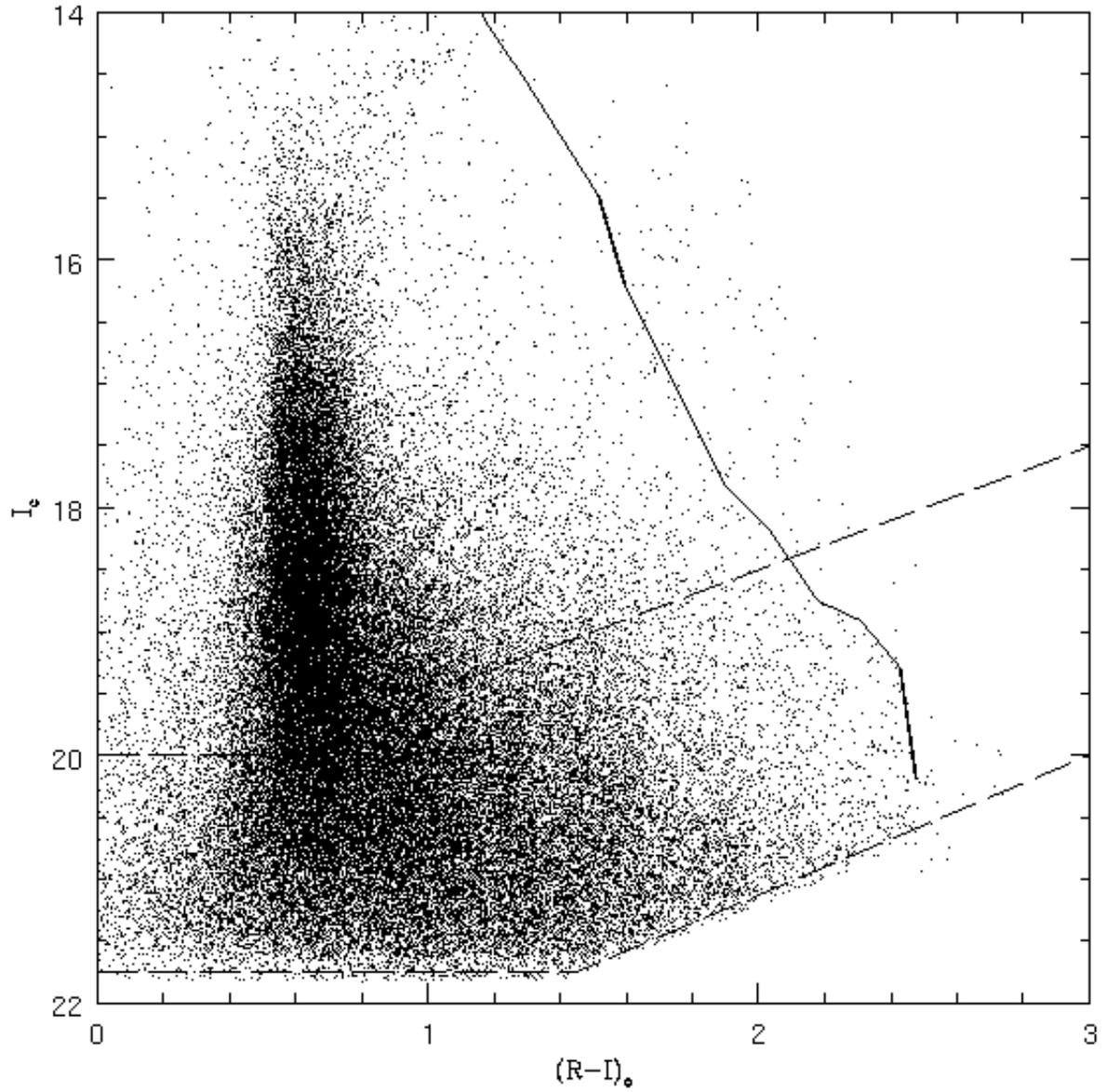}
\caption{All the photometry extracted from the 4m CTIO/BTC
survey. The solid line indicates the positions of a ZAMS, whereas the 
long-dashed lines show the location of the detection and completeness
limits.}
\end{figure*}

\setcounter{figure}{2}
\begin{figure*}
\vspace{18cm}
\includegraphics{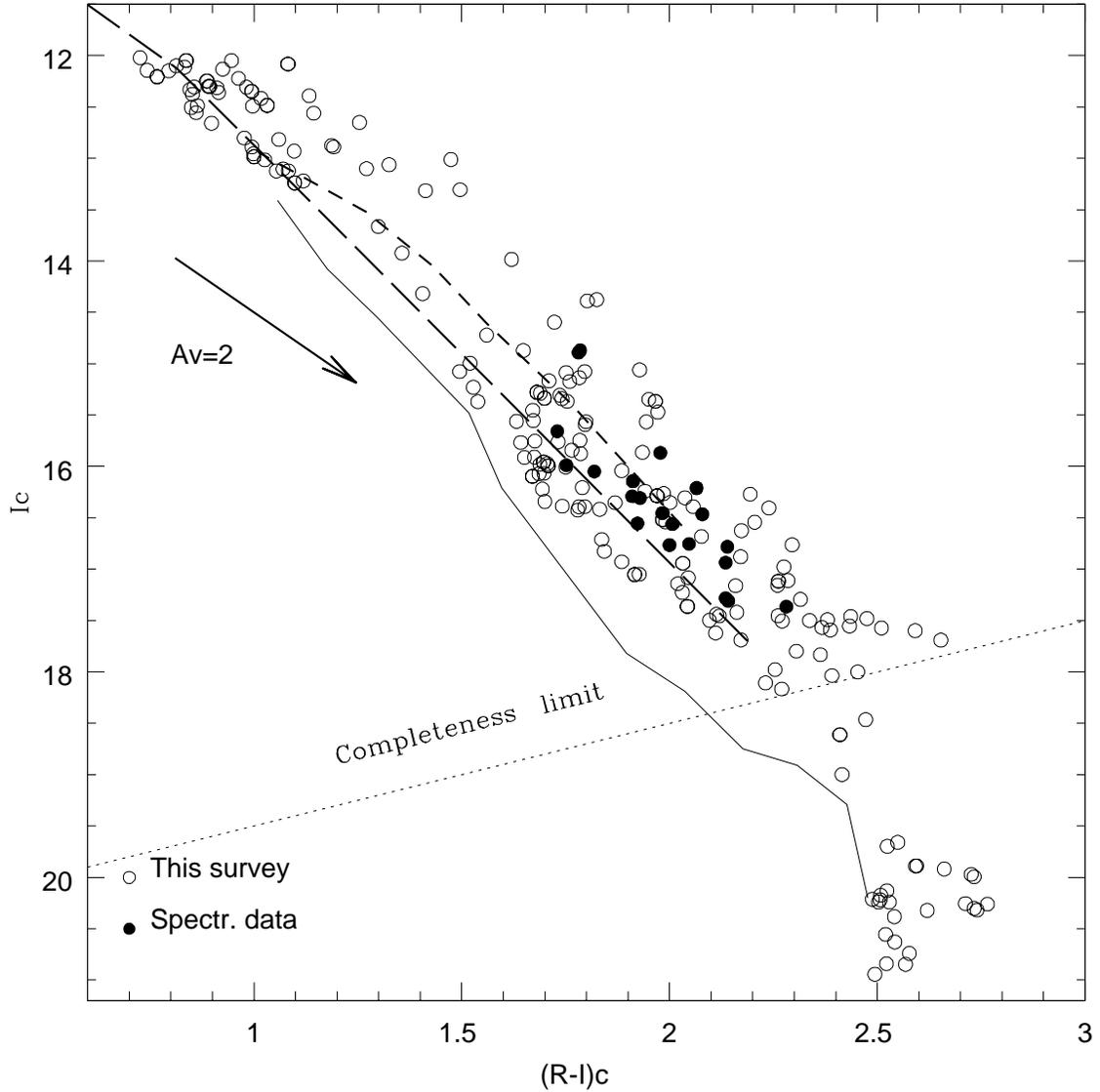}

\caption{Initial selection of candidates of IC2391 (open circles).
Spectroscopically confirmed members (Barrado y Navascu\'es et al. 1999)
 are displayed as solid circles. The solid, long-dashed and short-dashed
 lines indicate the positions of a ZAMS, an empirical IC2391 isochrone
and a 50 Myr isochrone (D'Antona \& Mazzitelli 1997). The completeness
limit of the BTC survey is indicated by a dotted line.}
\end{figure*}

\setcounter{figure}{3}
\begin{figure*}
\vspace{18cm}
\includegraphics{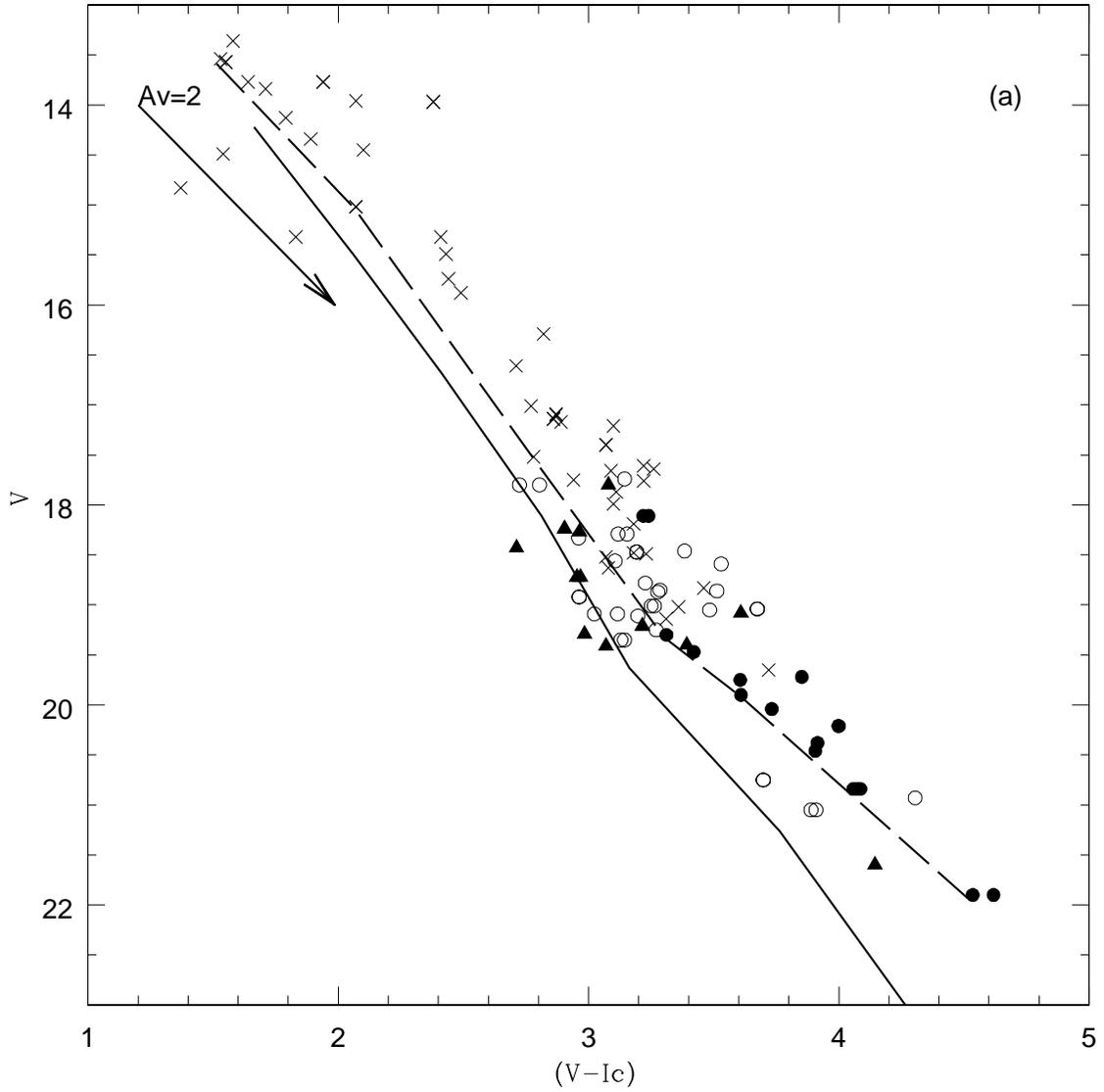}

\caption{{\bf a} Color-magnitude diagram of IC2391.  Solid circles represent 
the probable members of the cluster, whereas open circles correspond
to possible members.
Open and solid triangles  are objects initially selected as members,
whose membership has been rejected based of these CC and CM diagrams.
Data from previous surveys appear as  crosses.
The solid line represents the locus of an empirical  Zero Age
main-sequence, whereas the dashed line correspond to an empirical isochrone.}
\end{figure*}

\setcounter{figure}{3}
\begin{figure*}
\vspace{18cm}
\includegraphics{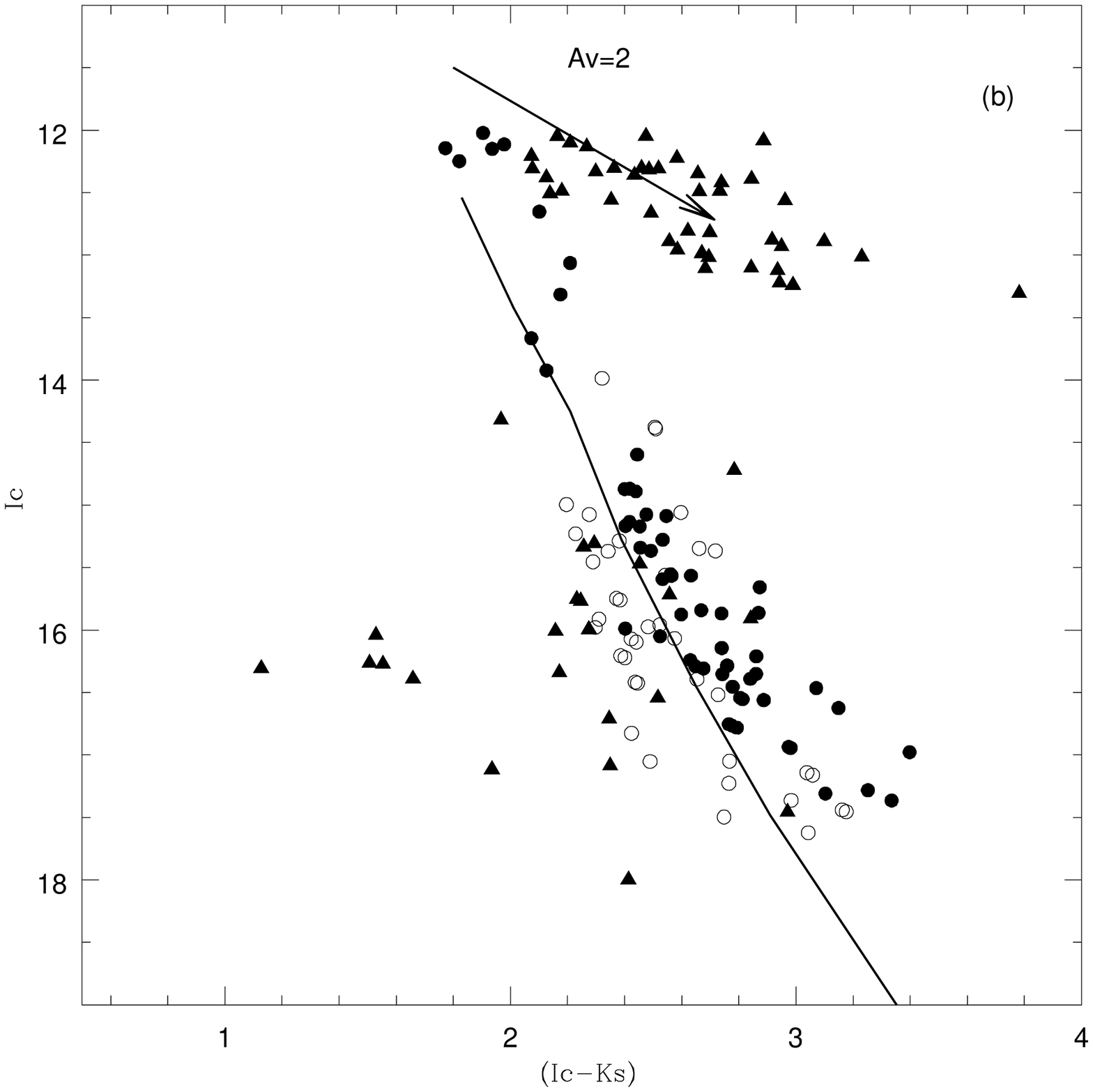}
\caption{{\bf b} Color-magnitude diagram of IC2391. Symbols as on Figure 4a.} 
\end{figure*}

\setcounter{figure}{3}
\begin{figure*}
\vspace{18cm}
\includegraphics{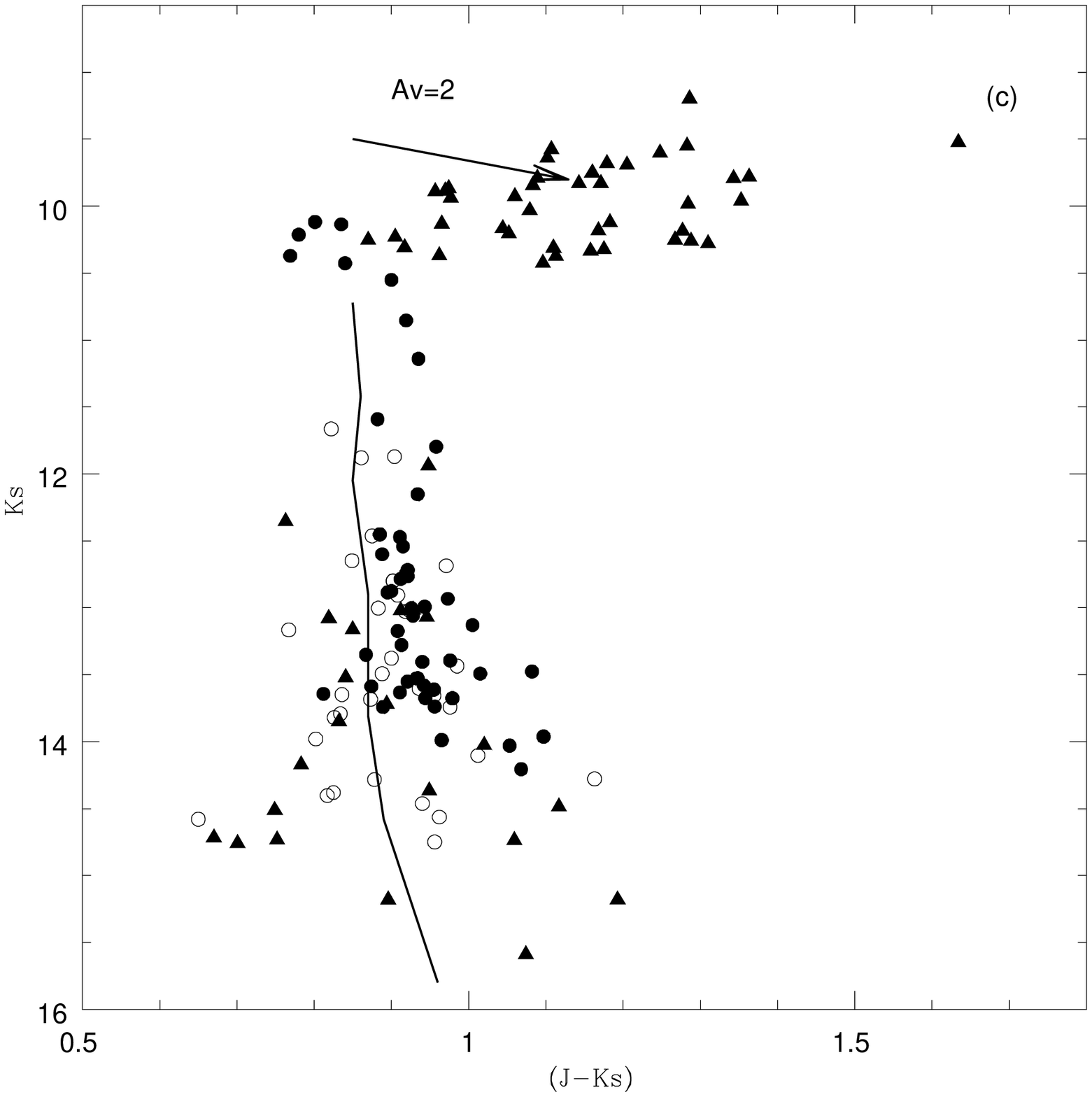}
\caption{{\bf c} Color-magnitude diagram of IC2391. Symbols as on Figure 4a.}
\end{figure*}

\setcounter{figure}{4}
\begin{figure*}
\vspace{18cm}
\includegraphics{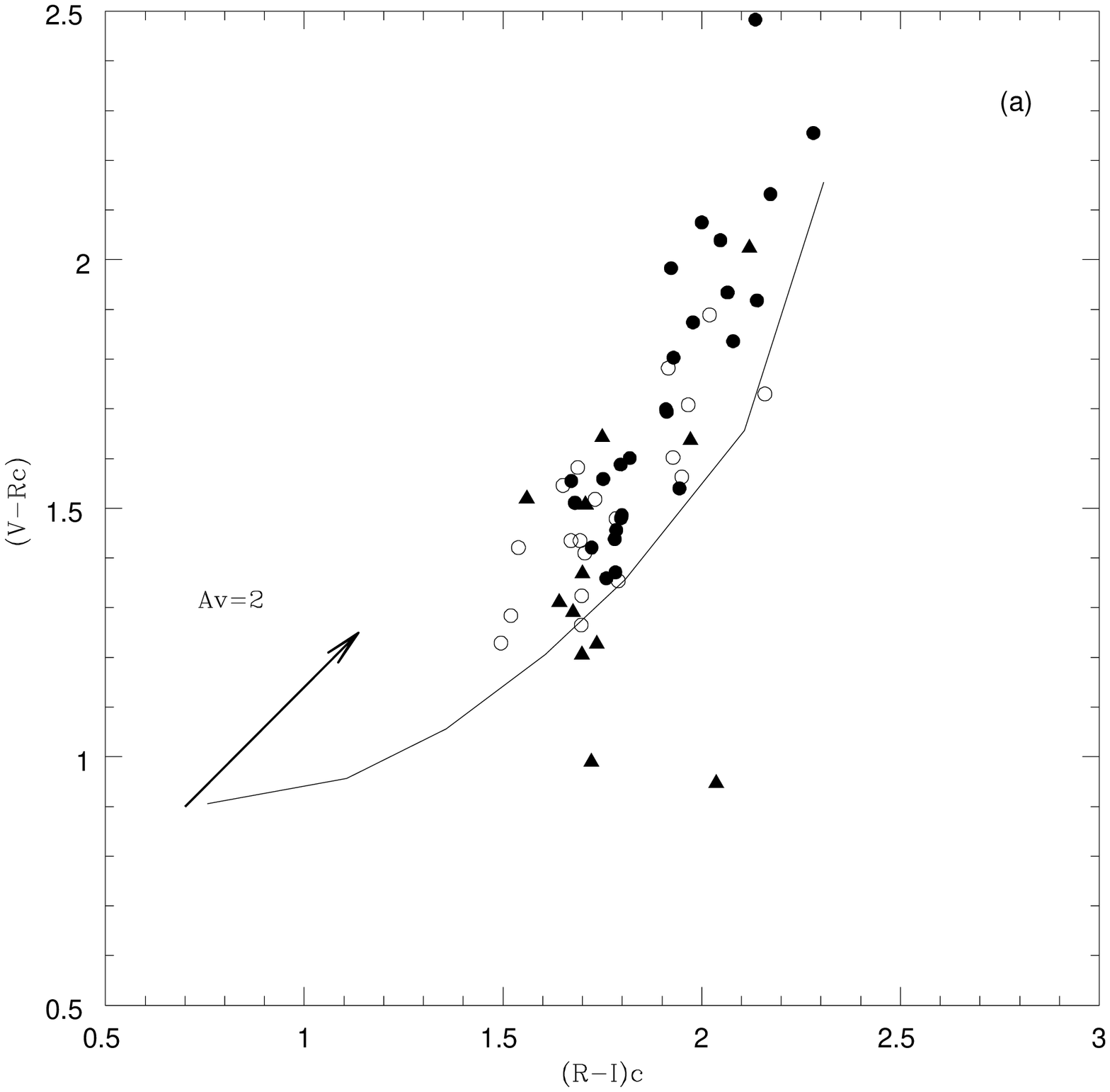}
\caption{{\bf a} Color-color diagram of IC2391.
Symbols as on Figure 4.}
\end{figure*}

\setcounter{figure}{4}
\begin{figure*}
\vspace{18cm}
\includegraphics{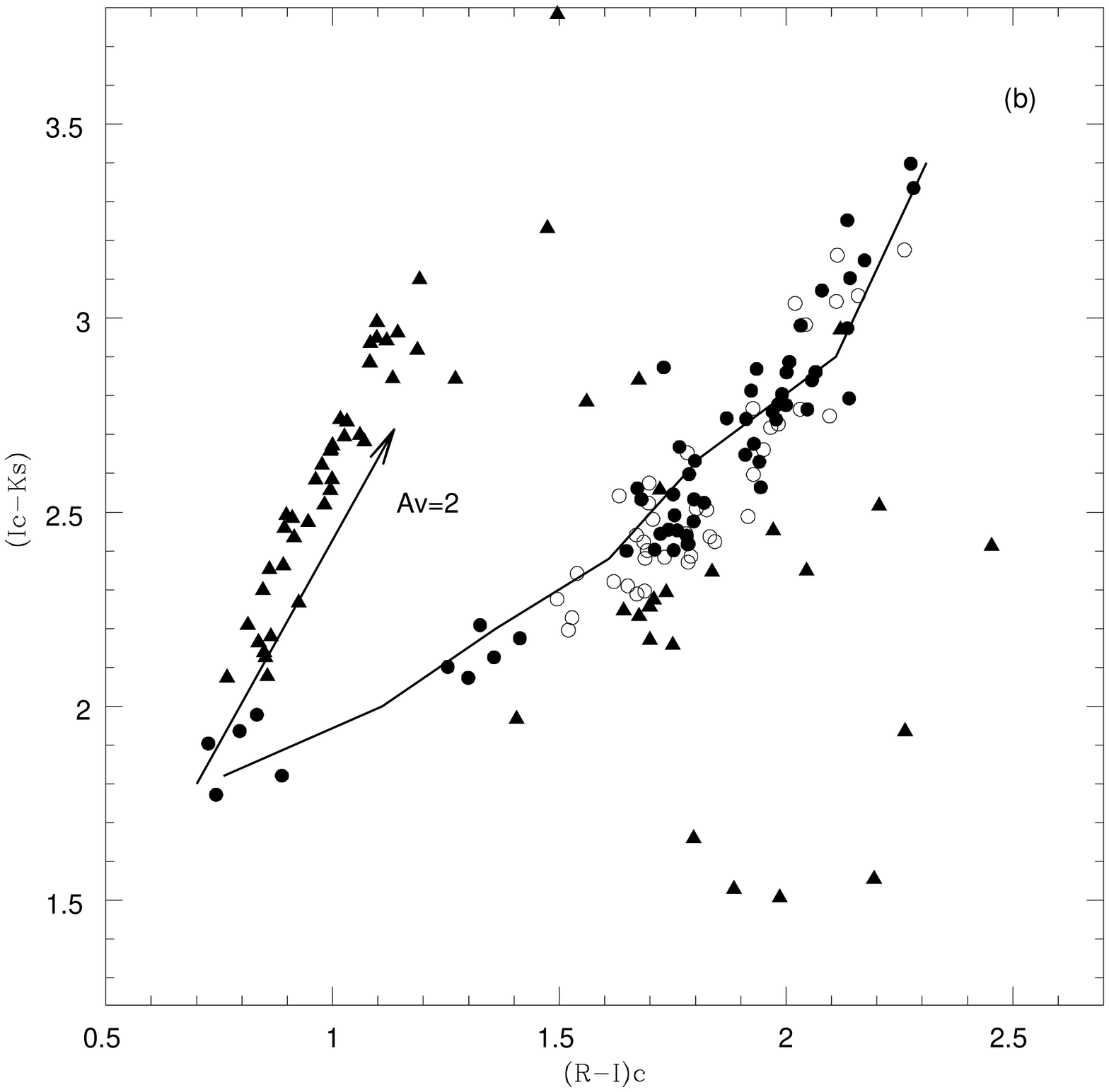}
\caption{{\bf b} Color-color diagram of IC2391.
Symbols as on Figure 4.}
\end{figure*}

\setcounter{figure}{5}
\begin{figure*}
\vspace{18cm}
\includegraphics{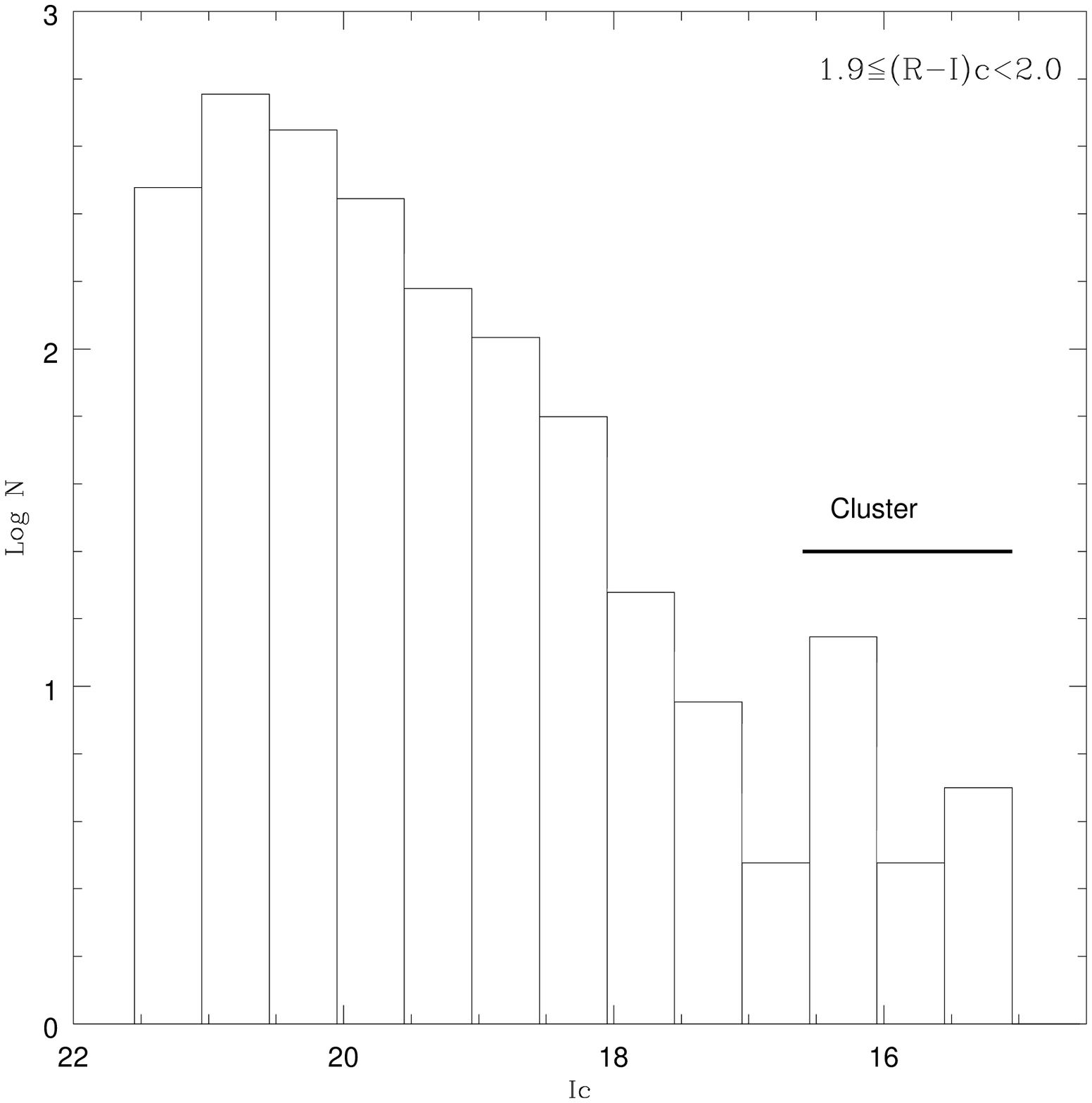}
\caption{Frequency of stars with 1.9$\le$(R-I)$_C$$<$2.0
against the I$_C$ magnitude. The location of IC2391 is indicated.}
\end{figure*}

\setcounter{figure}{6}
\begin{figure*}
\vspace{18cm}
\includegraphics{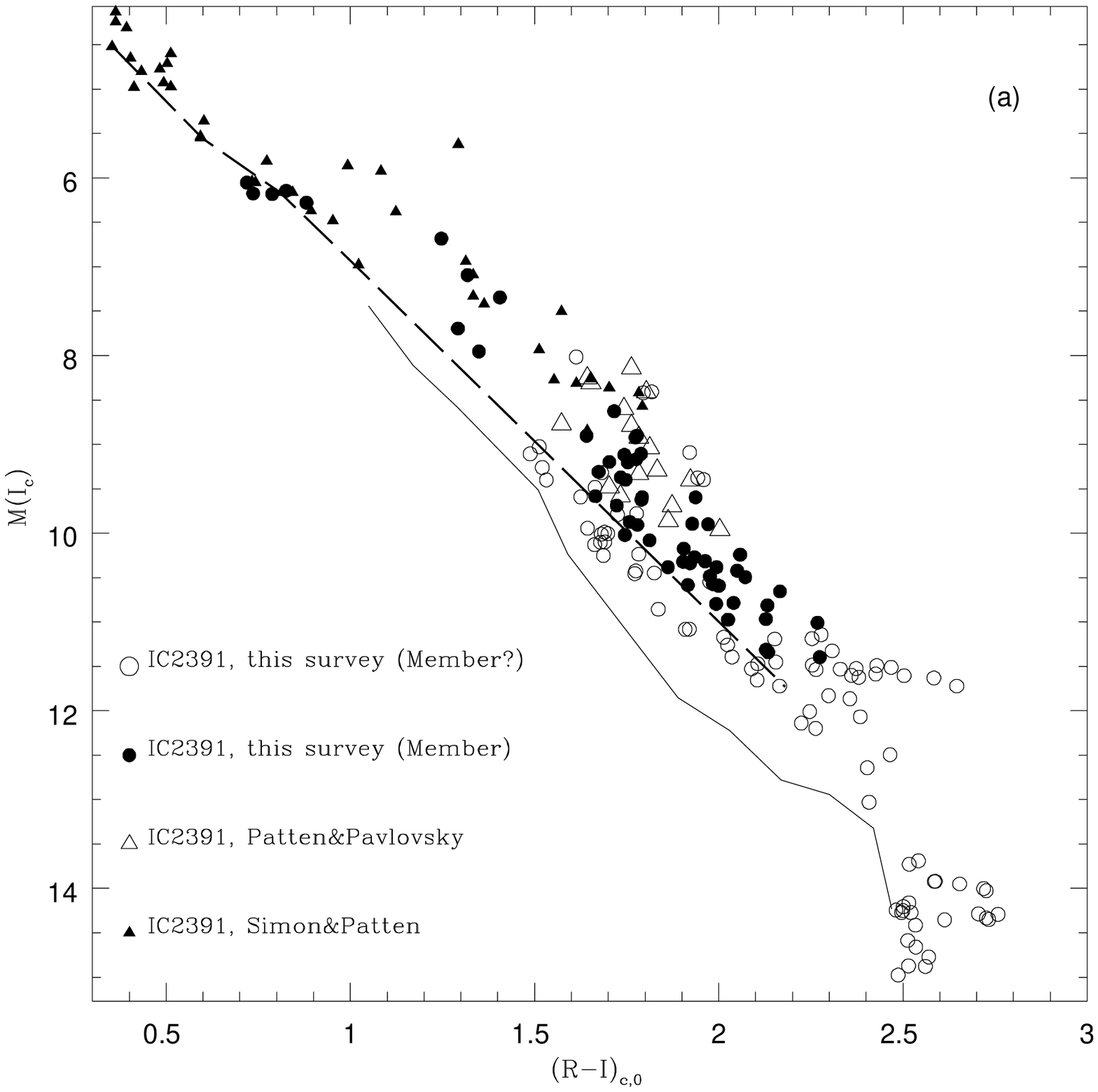}
\caption{{\bf a} Comparison with data from previous searches of
members of IC2391.}
\end{figure*}

\setcounter{figure}{6}
\begin{figure*}
\vspace{18cm}
\includegraphics{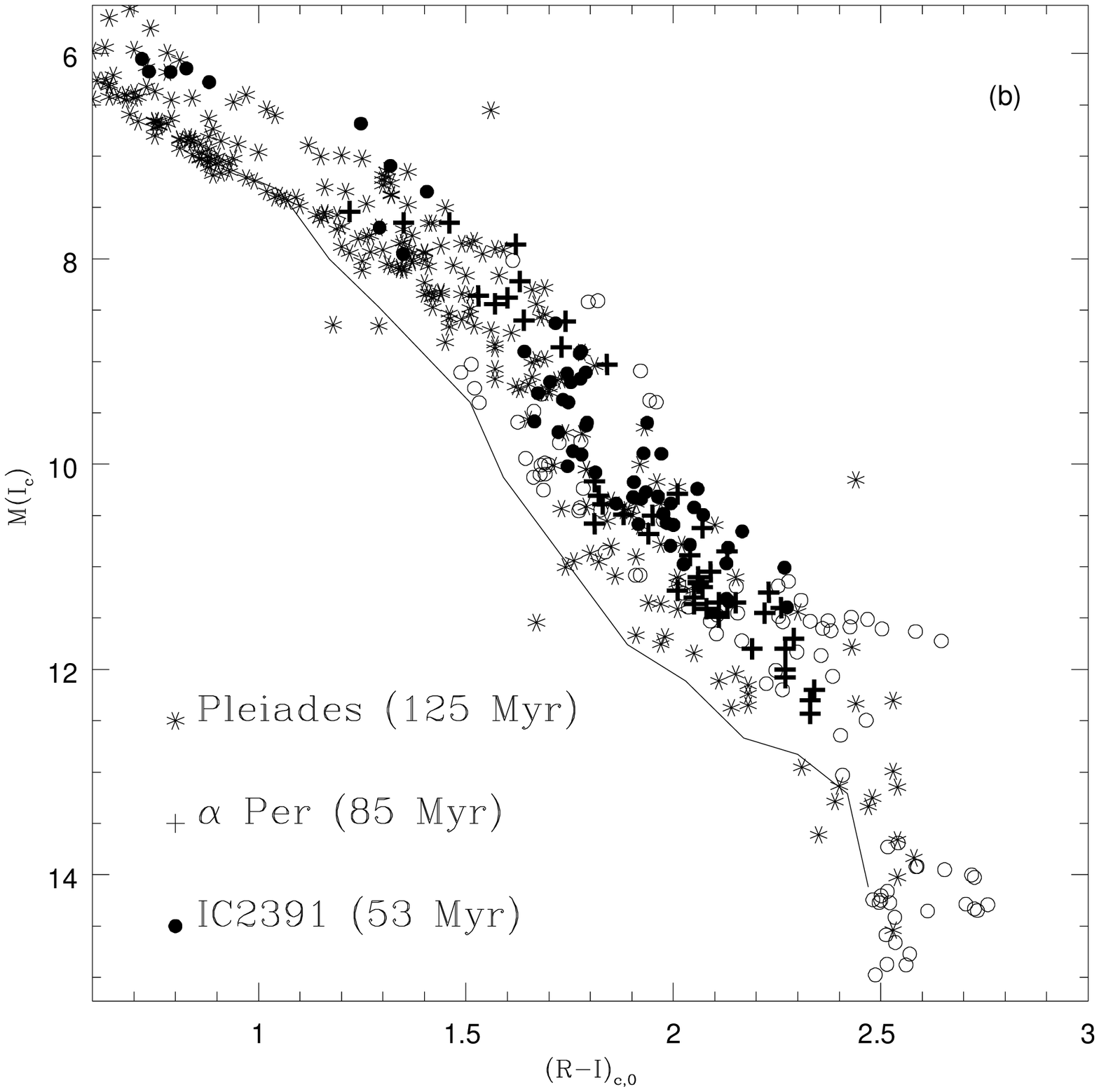}
\caption{{\bf b} Comparison between IC2301 candidate members (solid 
and open circles represent probable and possible members) and
members of Alpha Per (crosses) and the Pleiades (asterisks)}
\end{figure*}

\end{document}